\documentclass [10pt]{article}
\usepackage {graphicx}
\usepackage[american]{babel}
\usepackage[cp1251]{inputenc}  %% 1

\begin{document}

\title{Chaos and hyperchaos of geodesic flows on curved manifolds
corresponding to mechanically coupled rotators: Examples and numerical study
}

\author{S.P. Kuznetsov}

\maketitle
\begin{center}
\textit{Udmurt State University, Universitetskaya 1, Izhevsk, 426034, \\
Russian Federation}
\end{center}

\maketitle
\begin{center}
\textit{Kotel'nikov's Institute of Radio-Engineering and Electronics of RAS, \\
Saratov Branch, Zelenaya 38, Saratov, 410019, Russian Federation}
\end{center}

\begin{abstract}
A system of $N$ rotators is investigated with a constraint given by a
condition of vanishing sum of the cosines of the rotation angles. Equations
of the dynamics are formulated and results of numerical simulation for the
cases $N$=3, 4, and 5 are presented relating to the geodesic flows on a
two-dimensional, three-dimensional, and four-dimensional manifold,
respectively, in a compact region (due to the periodicity of the
configuration space in angular variables). It is shown that a system of
three rotators demonstrates chaos, characterized by one positive Lyapunov
exponent, and for systems of four and five elements there are, respectively,
two and three positive exponents (``hyperchaos''). An algorithm has been
implemented that allows calculating the sectional curvature of a manifold in
the course of numerical simulation of the dynamics at points of a
trajectory. In the case of $N$=3, curvature of the two-dimensional manifold is
negative (except for a finite number of points where it is zero), and
Anosov's geodesic flow is realized. For $N$=4 and 5, the computations show that
the condition of negative sectional curvature is not fulfilled. Also the
methodology is explained and applied for testing hyperbolicity based on
numerical analysis of the angles between the subspaces of small perturbation
vectors; in the case of $N$=3, the hyperbolicity is confirmed, and for $N$=4 and 5 the hyperbolicity does not take place.
\end{abstract}

Geodesic flows represent a special class of dynamical systems, the phase space of
which is the set of points of a certain manifold together with the velocity
vectors at these points \cite{1,2}. Motion takes place along geodesic lines, with
preservation of the velocity component in the direction of the geodesic.
This is a natural generalization of the free motion of a material point by
inertia in Euclidean space to the case of motion on a curved surface or in a
multidimensional space with curvature.

Geodesic flows on compact manifolds of negative curvature serve as a classic
example of deterministic chaos, which belongs to the category of hyperbolic
dynamics possessing the property of roughness or structural stability. Early
examples go back to Hadamard \cite{3}, but the valuable development of the
corresponding part of the theory of dynamical systems progressed in
1960s-1970s due to seminal work of Anosov and other researchers \cite{4o,4,5,6}.

Natural area of application of the geodesic flows with nontrivial dynamics
relates to mechanical systems that perform motion under conditions of
conservation of mechanical energy with imposed holonomic constraints being
defined by algebraic relations between the coordinates. The constraint
equations determine shape of the manifold as an object embedded in the
ambient space, and the metric on the manifold is naturally determined by the
quadratic form of the expression for kinetic energy through generalized
velocities.

Fig.~1a shows a mechanism built from three two-link cranks, which was
proposed as an example of a system with configuration space being a
nontrivial manifold by Thurston and Weeks \cite{7}. It can be interpreted as a
system of three rotators with a constraint defined by certain algebraic
relation between the angular coordinates. Hunt and MacKay showed that the
free motion of this mechanism with proper selection of parameters provides a
feasible example of the Anosov flow on compact two-dimensional manifold of
negative curvature \cite{8}. The special case, when the constraint equation is
simplified (in certain asymptotics in parameters) and reduces to a condition
of vanishing sum of cosines of the three angles of revolution of the
rotators about their fixed axes, was discussed in detail in \cite{9,10}. The
configuration space in this case is a two-dimensional curved surface, known
as the Schwarz P-surface \cite{11}. Also generalizations were considered that
imply replacement of the algebraic constraint equations by a potential
interaction of the rotators of such kind that the minimum of the potential
function takes place just on the Schwarz surface \cite{12,13,14}. Introduction of
dissipation and feedbacks into the system was discussed, which allowed
realizing chaotic self-oscillations generated by a hyperbolic structurally
stable attractor. On this base an electronic device reproducing the Anosov
dynamics was designed, which is a generator of rough chaos with
characteristics attractive for applications \cite{13,14}.

It seems natural to turn to generalizations consisting in considering
systems based on a different number of rotators.

A case of two rotators with constraint implemented by means of a hinge
mechanism appears to be not interesting in respect to its dynamics, as it is
rather trivial since the configuration space degenerates into a set of
intersecting straight lines on the plane of two angular variables. However,
by replacing the mechanical constraint by interaction of the rotators by
potential forces, some nontrivial motions such as wandering on a
two-dimensional lattice in the plane of angular variables can be
observed \cite{15}, similar to those orbits on the phase plane of non-autonomous systems, which provide
the so-called Zaslavsky stochastic web \cite{16}.

\begin{figure}[htbp]
\centerline{\includegraphics[width=5.5in]{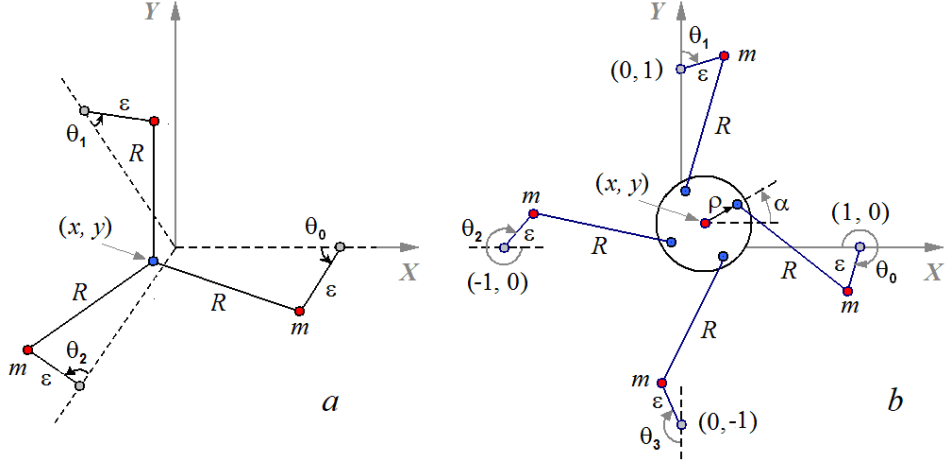}}
\label{fig1}
\caption{
Hinge mechanisms based on rotators, whose angles of revolution about
fixed axes are given by $\theta _i (t)$, with imposed holonomic constraint,
which in the limiting case $\varepsilon < < R = 1$ reduces to the equation
$\sum\nolimits_{i = 0}^{N - 1} {\cos \theta _i } = 0$, where $N$=3 (a) or 4
(b).
}
\end{figure}

The subject of this article is consideration of the systems constituted
by a larger number of rotators $N$ (concretely, the cases of $N$=3, 4, 5 will be
discussed), where the angles of rotation $\theta _0 ,\,\,\theta _1
,\,\,...\,,\,\theta _{N - 1} $ satisfy the constraint equation
\begin{equation}
\label{eq1}
\sum\limits_{i = 0}^{N - 1} {\cos \theta _i } = 0.
\end{equation}

In particular, for $N$=4, the corresponding situation can be implemented in the
hinge mechanism shown in Fig. 1b, under assumption of certain asymptotics in
parameters (see Appendix).

In Section 1, equations are formulated for a system of $N$ rotators with
constraint defined by the condition that sum of the cosines of the angles of
rotation equals zero. Two forms of equations are presented, one in the
ambient space with the imposed constraint condition, which is
convenient when numerically integrating the equations and processing the
data, and second is an equivalent formulation using the standard notation of
theory of Riemannian manifolds \cite{17}, which allows establishing and
discussing the relationship with the context of Anosov's theory. Section 2
presents and compares the numerical results of analysis of the dynamics for
$N$=3, which is the case of Anosov's dynamics corresponding to chaos
with one positive Lyapunov exponent, $N$=4 that is hyperchaotic dynamics with two
positive Lyapunov exponents on a three-dimensional curved manifold, and
$N$=5 that is hyperchaotic dynamics on a four-dimensional manifold with three
positive Lyapunov exponents. Section 3 introduces algorithm for calculating
the curvature at the points of the trajectory, which was tested by comparing
with the analytical formula for the case $N$=3, where the curvature is negative
except for a finite number of points on the manifold, and dynamics is
Anosov's. For the cases $N$=4 and 5 it is shown that the condition of negative
sectional curvature is not fulfilled: while moving along typical
trajectories, positive values of sectional curvatures also occur, so that
the geodesic flows under consideration cannot be classified as Anosov's
basing on the curvature. Section 4 discusses the technique and presents the
results of testing the hyperbolicity of the observed dynamics on a base of
analysis of statistics for the intersection angles for subspaces of vectors
of small perturbations. It is shown numerically that in the case of $N$=3 the
hyperbolicity is confirmed, and in the cases $N$=4 and 5 there is no
hyperbolicity. In conclusion, we discuss possible prospects for development
of the outlined approaches to analysis of chaotic dynamics, both for
geodetic flows and for some other types of nonlinear dynamical systems
implementing similar dynamics.

\section{Basic equations}
Consider a system composed of $N>2$ rotators characterized by time-dependent instantaneous angles of rotation
$\theta _i ,\,\,i = 0,\,\,1,\,\,2,\,...,\,N - 1$, each of which possesses unit moment
of inertia, assuming that the kinetic
energy is given by
\begin{equation}
\label{eq2}
T = \textstyle{1 \over 2}\sum\limits_{i = 0}^{N - 1} {\dot {\theta }_i^2 },
\end{equation}
and impose the constraint
\begin{equation}
\label{eq3}
F(\theta _0 ,\,\,\theta _1 ,...,\,\,\theta _{N - 1} ) = \sum\limits_{i =
0}^{N - 1} {\cos \theta _i } = 0.
\end{equation}

For $N$=3 and 4 it can be implemented by means of the hinge mechanisms
mentioned in the Introduction.

The Lagrange equations of motion of the system are
\begin{equation}
\label{eq4}
\ddot {\theta }_i = \Lambda \partial F / \partial \theta _i = - \Lambda \sin
\theta _i ,\,\,i = 0,\,\,1,\,\,N - 1,
\end{equation}
where the factor $\Lambda $ is to be determined taking into account the
algebraic mechanical constraint condition (\ref{eq3}) that is complementary to the
differential equations.

Differentiating the constraint equation twice
\begin{equation}
\label{eq5}
\,\sum\limits_{i = 0}^{N - 1} {\ddot {\theta }_i \sin \theta _i } +
\sum\limits_{i = 0}^{N - 1} {\dot {\theta }_i^2 \cos \theta _i } = 0,
\end{equation}
and substituting here the expressions for the second derivatives from (\ref{eq4}), we
obtain an explicit formula for the factor $\Lambda $
\begin{equation}
\label{eq6}
\Lambda = \frac{\sum\nolimits_{i = 0}^{N - 1} {\dot {\theta }_i^2 \cos
\theta _i } }{\sum\nolimits_{i = 0}^{N - 1} {\sin ^2\theta _i } }.
\end{equation}

Thus, we arrive at the following closed set of equations
\begin{equation}
\label{eq7}
\ddot {\theta }_i = -\frac{\sum\nolimits_{j = 0}^{N - 1} {\dot {\theta }_j^2
\cos \theta _j } }{\sum\nolimits_{j = 0}^{N - 1} {\sin ^2\theta _j } }\sin
\theta _i ,\,\,\,i = 0,\,1,\,...,\,N - 1,
\end{equation}
or, equivalently,
\begin{equation}
\label{eq8}
\dot {\theta }_i = u_i ,\,\,\,\dot {u}_i =- \frac{\sum\nolimits_{j = 0}^{N -
1} {u_j^2 \cos \theta _j } }{\sum\nolimits_{j = 0}^{N - 1} {\sin ^2\theta _j
} }\sin \theta _i ,\,\,\,i = 0,\,1,\,...,\,N - 1.
\end{equation}

The constraint condition (\ref{eq3}) and the relation obtained by its
differentiation correspond to two first integrals of this system, which
formally has order 2$N$. The initial conditions must be selected with
restriction that these first integrals have the required values, namely,
\begin{equation}
\label{eq9}
\sum\limits_{i = 0}^{N - 1} {\cos \theta _i } = 0
\end{equation}
and
\begin{equation}
\label{eq10}
\sum\limits_{i = 0}^{N - 1} {u_i \sin \theta _i } = 0.
\end{equation}

Dynamics of the system (\ref{eq8}) can be interpreted as motion of a particle on
manifold of dimension $N - 1$ defined by equation (\ref{eq3}), along geodesic
lines of the metric provided by quadratic form corresponding to the
expression for the kinetic energy, namely
\begin{equation}
\label{eq11}
ds^2 = \sum\limits_{i = 0}^{N - 1} {d\theta _i^2 } ,
\end{equation}
with the condition $\sum\limits_{i = 0}^{N - 1} {\sin \theta _i } \,d\theta
_i = 0$ arising due to the constraint equation. Since the quantities $\theta
_i $ have the meaning of angular coordinates, they can be considered as
related to intervals $(-\pi,\,\pi)$, and the dynamics may be regarded as
occurring in a compact domain.

As evident from the equations, the motions that differ only by the value of
the kinetic energy are identical up to a time scale and a value of velocity along
the geodesic. In this regard, all the specific data will be given hereafter only for
the case of a unit absolute value of velocity, which corresponds to the
energy $T = \textstyle{1 \over 2}$.

The dynamics of the systems for the number of rotators 3, 4, and 5 are
illustrated in Fig. 2 by graphs of angular velocities as functions of time
obtained from numerical integrating equations (\ref{eq8}) by means of the Runge --
Kutta method. As seen from the figure, in all cases the behavior looks
chaotic: the dependences demonstrate obvious irregularity and absence of
visible repetition of forms. The chaotic nature of the dynamics is also
confirmed by the type of power spectra shown in Fig.3. The spectra were
obtained by processing the time series for angular velocity of one of the
rotators using data of numerical integration of the equations, with
application of the method of statistical evaluation of the spectral density
developed for random processes \cite{18, 19}. For this, the time series is
divided into sections of certain duration, significantly exceeding the
characteristic time scale of the signal, with multiplying data for each
segment by the ``window'' function (to improve the quality of spectral
analysis due to removing, as far as possible, the effect of signal mismatch at the edges of the
partition intervals). Next, the Fourier transform is performed for each segment,
and the squares of the amplitudes of the spectral components are averaged
over a set of the segments. It can be seen from the figure that in each of
the cases considered the spectra are continuous, like for random processes;
there are no expressed discrete components. The chaotic nature of the
dynamics is also confirmed by analysis of Lyapunov exponents in the next
Section.

\begin{figure}[htbp]
\centerline{\includegraphics[width=5in]{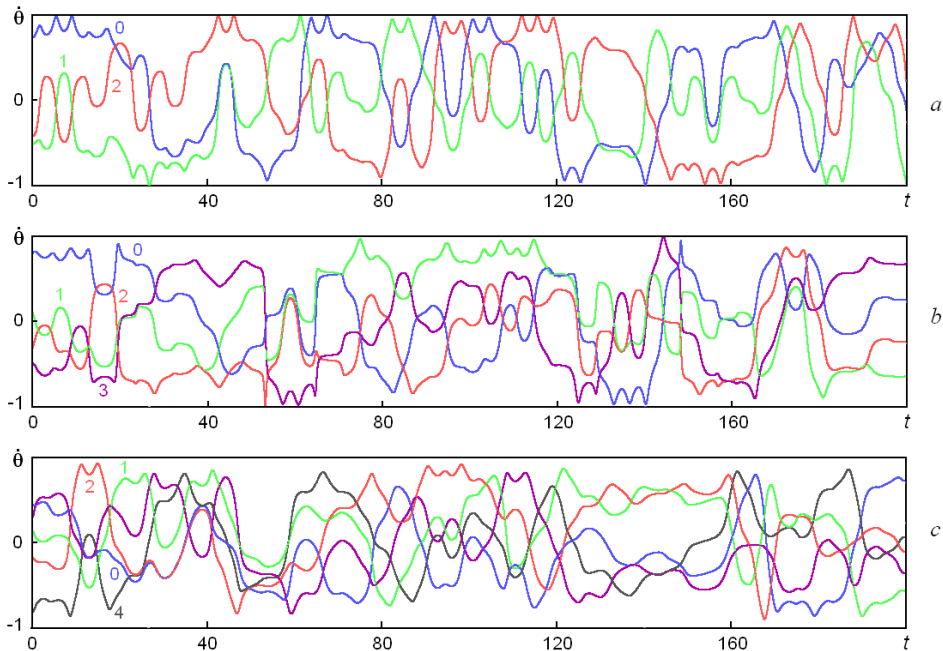}}
\label{fig2}
\caption{
Waveforms of the angular velocities for the system (\ref{eq8}) in the cases
$N$=3 (a), $N$=4 (b) and $N$=5 (c) obtained numerically for motions with unit
velocity in the direction of the geodesics.
}
\end{figure}

\begin{figure}[htbp]
\centerline{\includegraphics[width=3.5in]{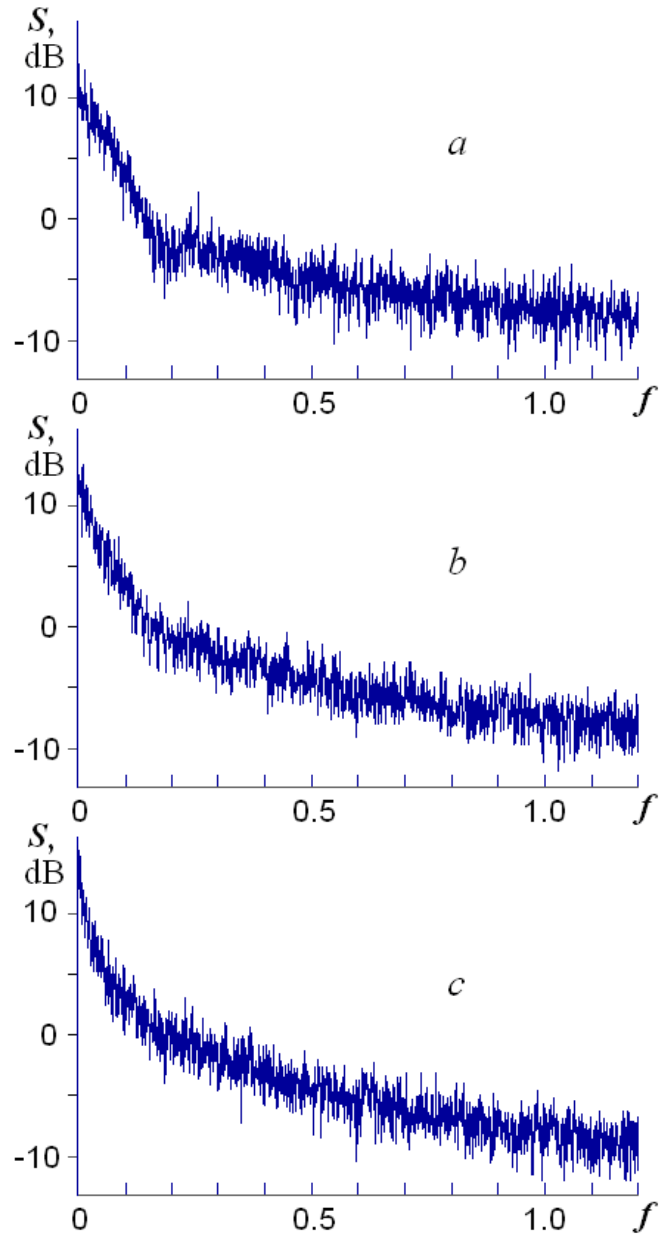}}
\label{fig3}
\caption{
Spectral power density plots of signal given by angular velocity of
one of the rotators as obtained from processing data of numerical simulation
of the system (\ref{eq8}) for the cases $N$=3 (a), $N$=4 (b) and $N$=5 (c) for the motion
with unit velocity along the geodesics.
}
\end{figure}
The above dynamical equations (\ref{eq7}) or (\ref{eq8}) can be rewritten using notation
adopted in the theory of Riemannian manifolds. For this, we express angular
coordinate and velocity of one of the rotators, say, $\theta _0 $ and $u_0
$, in terms of coordinates and velocities of the other rotators, while the
set of their angular velocities $(u_1 ,\,u_2 ,...,\,u_{N - 1} )$ is assumed
to be components of a contravariant vector $u^\alpha ,\,\,\alpha =
1,\,2,\,\,...\,\,,\,N - 1$. Substituting the result into the formula for
kinetic energy, we have
\begin{equation}
\label{eq12}
T = \textstyle{1 \over 2}g_{\alpha \beta } u^\alpha u^\beta ,
\end{equation}
where the Greek indices run from 1 to $N - 1$, and summation is assumed over
repeating superscripts and superscripts. Components of the metric tensor
$g_{\alpha \beta } $ are given by the relations
\begin{equation}
\label{eq13}
g_{\alpha \beta } = \delta _{\alpha \beta } + \frac{\sin \theta _\alpha \sin
\theta _\beta }{1 - \left( {\sum\nolimits_{\gamma = 1}^{N - 1} {\cos \theta
_\gamma } } \right)^2}.
\end{equation}

The equation of motion then takes the form known for geodesic flows
\cite{1,2,4,5}, namely,
\begin{equation}
\label{eq14}
\dot {u}^\alpha + \Gamma _{\beta \gamma }^\alpha u^\beta u^\gamma = 0,
\end{equation}
where the Christoffel symbols $\Gamma _{\beta \gamma }^\alpha $ are
involved, which have expressions through the components of the metric tensor
given by standard formulas, and in our particular case they are
\begin{equation}
\label{eq15}
\Gamma _{\beta \gamma }^\alpha = \frac{\delta _{\beta \gamma } \cos \theta
_\beta - \sin \theta _\beta \sin \theta _\gamma \sum\nolimits_{\nu = 1}^{N -
1} {\cos \theta _\nu } \left[ {1 - \left( {\sum\nolimits_{\nu = 1}^{N - 1}
{\cos \theta _\nu } } \right)^2} \right]^{ - 1}}{1 + \sum\nolimits_{\nu =
1}^{N - 1} {\sin ^2\theta _\nu } - \left( {\sum\nolimits_{\nu = 1}^{N - 1}
{\cos \theta _\nu } } \right)^2}\sin \theta _\alpha .
\end{equation}

Among formulas of Riemannian geometry, it is useful to keep in mind the
relation that allows tracing components of a vector $p^\alpha $ in the
course of parallel transport along a geodesic:
\begin{equation}
\label{eq16}
\dot {p}^\alpha + \Gamma _{\beta \gamma }^\alpha u^\beta p^\gamma = 0.
\end{equation}

The above relations, where $\theta _0 $ and $u^0$ play the distinguished
role, correspond to a particular choice of one of $N$ charts constituting the
atlas of the manifold under consideration. For other charts of the atlas the
formulas are similar and are obtained by cyclic permutations of the indices.
%---------------------------------
\section{Lyapunov exponents}
Formally, the dimension of the phase space of the system (\ref{eq8}) is 2$N$.
Respectively, there are 2$N$ Lyapunov exponents characterizing the behavior of
disturbances near a reference trajectory. However, among them, there are two
nonphysical zero exponents, which correspond to
perturbations of the integrals of motion responsible for the imposed
constraint, namely, the perturbations violating the constraint
equations. They should be excluded; therefore, the total number of
significant Lyapunov exponents is $2N - 2$. Due to the conservative nature
of the system and symmetry with respect to the time reversal, the presence
of each positive exponent implies the presence of another negative one, with
equal absolute value. The spectrum of relevant Lyapunov
exponents also contains necessarily two zero exponents, one of which appears
because of the autonomous nature of the system being responsible for a
perturbation directed tangentially to the phase trajectory, and the second
is associated with a perturbation produced by an energy shift.

From the above, it follows that to obtain the full spectrum of Lyapunov
exponents it is sufficient to calculate the largest $N - 2$ exponents. This
can be done within the framework of computational procedure that implements
the joint solution of the dynamical equations (\ref{eq8}) together with a set of
$N - 2$ equations in variations \cite{20,21,22,23,24}. The equations in variations are
obtained by substituting in (\ref{eq8}) dynamic variables with added
perturbations: $\theta _i \to \theta _i + \tilde {\theta }_i ,\,\,u_i \to
u_i + \tilde {u}_i $, and by linearization retaining the first-order
terms for small perturbations:
\begin{equation}
\label{eq17}
\left\{ {\begin{array}{l}
 \dot {\tilde {\theta }}_i = \tilde {u}_i , \\
 \\
 {\dot {\tilde {u}}_i = -\frac{\sum\nolimits_{j = 0}^{N - 1} {(2u_j
\tilde {u}_j \cos \theta _j - u_j^2 \tilde {\theta }_j \sin \theta _j )}
}{\sum\nolimits_{j = 0}^{N - 1} {\sin ^2\theta _j } }\sin \theta _i - \left(
{\frac{\sum\nolimits_{j = 0}^{N - 1} {u_j^2 \cos \theta _j }
}{\sum\nolimits_{j = 0}^{N - 1} {\sin ^2\theta _j } }} \right)\tilde {\theta
}_i \cos \theta _i } \\
 { + \frac{\left( {\sum\nolimits_{j = 0}^{N - 1} {u_j^2 \cos \theta _j } }
\right)}{\left( {\sum\nolimits_{j = 0}^{N - 1} {\sin ^2\theta _j } }
\right)^2}\sum\nolimits_{j = 0}^{N - 1} {\tilde {\theta }_j \sin 2\theta _j
} ,\,\,\,i = 0,\,1,\,...,\,N -1.
} \\
\end{array}} \right.
\end{equation}

In the framework of the computational procedure \cite{20,21,22,23,24}, at each step of
integrating the differential equations, the obtained $N - 2$ perturbation
vectors ${\rm {\bf \xi }} = (\tilde {\theta }_0 ,\,...,\,\tilde {\theta }_{N
- 1} ,\,\tilde {u}_0 ,\,...,\,\tilde {u}_{N - 1} )$ are subjected to Gram --
Schmidt orthogonalization and normalization:
\begin{equation}
\label{eq18}
\begin{array}{l}
 {\rm {\bf x}}_1 = {\rm {\bf \xi }}^{(\ref{eq1})}(t),\,\,\, \\
 {\rm {\bf \tilde {x}}}^{(\ref{eq1})}(t) = {\rm {\bf x}}_1 / \left\langle {{\rm {\bf
x}}_1 ,\,{\rm {\bf x}}_1 } \right\rangle , \\
 {\rm {\bf x}}_2 = {\rm {\bf \xi }}^{(2)}(t) - \left\langle {{\rm {\bf \xi
}}^{(2)}(t),{\rm {\bf \tilde {x}}}^{(\ref{eq1})}(t)} \right\rangle {\rm {\bf \tilde
{x}}}^{(\ref{eq1})}(t),\,\,\, \\
 {\rm {\bf \tilde {x}}}^{(2)}(t) = {{\rm {\bf x}}_2 } \mathord{\left/
{\vphantom {{{\rm {\bf x}}_2 } {\left\langle {{\rm {\bf x}}_2 ,\,{\rm {\bf
x}}_2 } \right\rangle }}} \right. \kern-\nulldelimiterspace} {\left\langle
{{\rm {\bf x}}_2 ,\,{\rm {\bf x}}_2 } \right\rangle },\,\,\, \\
 {\rm {\bf x}}_3 = {\rm {\bf \xi }}^{(3)}(t) - \left\langle {{\rm {\bf \xi
}}^{(3)}(t),{\rm {\bf \tilde {x}}}^{(\ref{eq1})}(t)} \right\rangle {\rm {\bf \tilde
{x}}}^{(\ref{eq1})}(t) - \left\langle {{\rm {\bf \xi }}^{(3)}(t),{\rm {\bf \tilde
{x}}}^{(2)}(t)} \right\rangle {\rm {\bf \tilde {x}}}^{(2)}(t),\,\, \\
 \,{\rm {\bf \tilde {x}}}^{(3)}(t) = {{\rm {\bf x}}_3 } \mathord{\left/
{\vphantom {{{\rm {\bf x}}_3 } {\left\langle {{\rm {\bf x}}_2 ,\,{\rm {\bf
x}}_2 } \right\rangle }}} \right. \kern-\nulldelimiterspace} {\left\langle
{{\rm {\bf x}}_2 ,\,{\rm {\bf x}}_2 } \right\rangle },\, \\
 \cdots \,\, \\
 \end{array}
\end{equation}
where the angle brackets denote the scalar product. In our case, it is
convenient to use the definition of the scalar product involving only the
coordinate components of the vectors, namely, $\left\langle {{\rm {\bf \xi
}}^{(s)},\,\,{\rm {\bf \xi }}^{(r)}} \right\rangle = \sum\limits_{j = 0}^{N
- 1} {\tilde {\theta }_j^{(s)} \tilde {\theta }_j^{(r)} } $. (This will be
useful and adequate also in the calculating the curvatures, as described in
the next Section.) A numerical estimate of the Lyapunov exponents is given
by cumulative sums of logarithms of norms $\sqrt {\left\langle {{\rm {\bf
x}}_i ,\,{\rm {\bf x}}_i } \right\rangle } $ divided by the observation
time.

Since there is no characteristic time scale in the system, the non-zero
Lyapunov exponents characterizing the exponential growth and decrease of
perturbations per unit of time are directly proportional to the velocity,
that is, to the square root of kinetic energy. In this regard, all results
are given here only for the case of unit velocities along the geodesics.

Numerical calculations performed for the cases $N$=3, 4, and 5 give the
following values for the Lyapunov exponents.

$N$=3:

\[
\lambda _1 = 0.500,\,\,\lambda _2 = 0,\,\,\lambda _3 = 0,\,\,\lambda _4 = -
0.500.
\]

$N$=4:

\[
\lambda _1 = 0.385,\,\,\lambda _2 = 0.230,\,\,\lambda _3 = 0,\,\,\lambda _4
= 0,\,\,\lambda _5 = - 0.230,\,\,\lambda _6 = - 0.385.
\]

$N$=5:

\[
\begin{array}{c}
\lambda _1 = 0.298,\,\,\lambda _2 = 0.227,\,\,\lambda _3 = 0.121,\,\,\lambda _4 = 0,\,\,\lambda _5 = 0,\\ \\
\lambda _6 = - 0.121,\,\,\lambda _7 = -0.227,\,\,\lambda _8 = - 0.298.
\end{array}
\]

As seen, in all three cases there are positive Lyapunov exponents that
indicates chaotic nature of the dynamics.

In the first case, $N$=3, when the motion takes place on a two-dimensional
manifold representing the Schwarz surface, there is one positive Lyapunov
exponent. In the second case, $N$=4, the motion takes place on a
three-dimensional curved manifold being characterized by two positive Lyapunov
exponents. Finally, in the case $N$=5, we have the motion on a four-dimensional
curved manifold, and there are three positive exponents. Chaotic behavior
with a number of positive Lyapunov exponents more than one, in
the literature is called hyperchaos \cite{25,26}.

%---------------------------------

\section{Curvature}
We now turn to consideration of a numerical method for evaluation of the
curvature of manifolds at points visited by trajectories. A complete
description of the local curvature of manifolds is given by the Riemann
tensor \cite{17}. However, bearing in mind the content of the Anosov theory, it
is natural to turn to an alternative method for quantifying the curvature,
namely, in terms of sectional curvatures \cite{4,5}. According to Anosov, the
hyperbolic nature of the chaotic dynamics of a geodesic flow and,
accordingly, its structural stability (roughness), is guaranteed if the
sectional curvature is everywhere negative.

Let us consider motion with a unit absolute velocity along a certain
reference trajectory representative for our geodesic flow, as well as motions obtained
by infinitesimal displacements in the directions orthogonal to the geodesic
with the same absolute velocity. Moving along the reference trajectory in
the process of numerical integrating the equations (\ref{eq8}), we simultaneously
integrate a set of equations in variations (\ref{eq17}), like in the procedure of
calculating the Lyapunov exponents. For each point $A \quad (\theta _0 ,\,\theta _1
,\,...,\,\theta _{N - 1} )$, visited when performing the numerical
integration of the equations, we define an orthonormal basis of
$N$ vectors ${\rm {\bf p}}^{(0)},\,...,\,\,{\rm {\bf p}}^{(N - 1)}$ in the
tangent space. It is assumed that the vector \textbf{p}$_{0}$ is directed
along the reference trajectory, i. e. ${\rm {\bf p}}^{(0)} = (u_0 ,\,u_1
,\,...,\,u_{N - 1} )$, and the rest ones are specified as the coordinate
components of the Lyapunov vectors at $A$, subjected to orthogonalization and
normalization, namely, ${\rm {\bf p}}^{(i)} = (\tilde {\theta }_0^{(i)}
,\,...,\,\tilde {\theta }_{N - 1}^{(i)} ),\,\,i = 1,\,...,\,N - 2$.

Further, applying the same difference scheme as used for integrating the
differential equations (\ref{eq8}) and the variational equations (\ref{eq17}), the
components of the reference basis vectors are calculated by their parallel
transfer along the reference geodesic at points shifted from the initial
position by one time step $\Delta t$ back and forth. This is done by
numerical solving the equations
\begin{equation}
\label{eq19}
\dot {p}_i^{(k)} = -\frac{\sum\nolimits_{j = 0}^{N - 1} {p_j^{(k)} u_j \cos
\theta _j } }{\sum\nolimits_{j = 0}^{N - 1} {\sin ^2\theta _j } }\sin \theta
_i ,\,\,\,i = 0,\,1,\,...,\,N - 1.
\end{equation}

As can be shown, these equations
correspond exactly to the relation (\ref{eq16}) in the notation of Riemannian geometry.

Finally, the coefficients $c^{(k)}$ for the expansion in the basis ${\rm
{\bf p}}^{(1)},\,...,\,\,{\rm {\bf p}}^{(N - 1)}$ are calculated for
Lyapunov vectors at the points $t\pm \Delta t$. These coefficients must
satisfy the following equation \cite{5}, p.136:

\begin{equation}
\label{eq20}
\frac{d^2c_i^{(k)} }{dt^2} + \sum\limits_{k = 1}^{N - 1} {K_{ij}
(t)c_j^{(k)} } = 0,
\end{equation}
where the symmetric matrix $K_{ij}$ of size $(N - 2)\times (N - 2)$ results
from convolution of the curvature tensor \textbf{R} at the analyzed point on
the manifold, $K_{ij} = \left\langle {{\rm {\bf R}}({\rm {\bf p}}^{(0)},{\rm
{\bf p}}^{(j)}),\,{\rm {\bf p}}^{(i)}} \right\rangle $.

Let ${\rm {\bf r}}$ be an arbitrary unit vector orthogonal to a vector ${\rm
{\bf p}}^{(0)}$ (directed along the geodesic). Then the value of the
quadratic form $\sum {K_{ij} } r_i r_j $ is the curvature of the manifold at
the point $A$ in a two-dimensional direction, defined by the vectors ${\rm {\bf
p}}^{(0)}$ and ${\rm {\bf r}}$, which is called the sectional curvature. In
particular, in the framework of the construction carried out, the matrix
element $K_{11}$ corresponds to the sectional curvature in the direction
given by the velocity vector and the first Lyapunov vector (associated with
the largest Lyapunov exponent), and the remaining diagonal elements
correspond to the curvatures in the directions given by the velocity vector
and the other Lyapunov vectors.

Having coefficients $c_i^{(k)} $ related to time points $t - \Delta
t,\,\,t,\,\,t + \Delta t$ and taking into account that $c_i^{(k)} (t) = \delta _{ik} $, one can use difference approximation for the second
derivative in (\ref{eq20}) and obtain the elements of the matrix $K_{ij}$ at $A$ numerically:
\begin{equation}
\label{eq21}
K_{ij} (t) = \frac{c_i^{(j)} (t - \Delta t) - 2c_i^{(j)} (t) + c_i^{(j)} (t
+ \Delta t)}{\Delta t^2}.
\end{equation}

In the case when the system of three rotators is considered, instead of the
matrix, we have a single value $K=K_{11}$, which is a function of angular
variables and represents the Gaussian curvature of the two-dimensional
manifold that is the Schwarz surface, given by the equation $\cos \theta _0
+ \cos \theta _1 + \cos \theta _2 = 0$. This curvature is expressed
explicitly \cite{8,9,10}:
\begin{equation}
\label{eq22}
K = - \frac{\cos ^2\theta _0 + \cos ^2\theta _1 + \cos ^2\theta _2 }{2(\sin
^2\theta _0 + \sin ^2\theta _1 + \sin ^2\theta _2 )^2}.
\end{equation}

With the analytical formula it is possible to test reliably
the above described algorithm by direct comparison of the
numerical results and the curvature values from (\ref{eq22}). The correspondence turns
out to be very good, for example, when specifying the integration step of
the fourth-order Runge -- Kutta method as $\Delta t = 0.01$, a coincidence
occurs up to the fifth significant digit.

Analysis of the formula (\ref{eq22}) shows that the curvature is negative over the
entire surface with exception of eight points $\theta _0 = \pm \pi /
2,\,\,\,\theta _1 = \pm \pi / 2,\,\,\,\theta _2 = \pm \pi / 2$, where it is
zero because of vanishing of all three cosines. The geodesic flow turns out
to belong to the class of Anosov systems, realizing hyperbolic chaos in its
conservative version. The presence of points where the curvature is zero
does not prevent this, since there are only a finite number of them in the
considered compact part of the manifold (the range of variation of the
angles from $-\pi$ to $\pi$).

\begin{figure}[htbp]
\centerline{\includegraphics[width=5in]{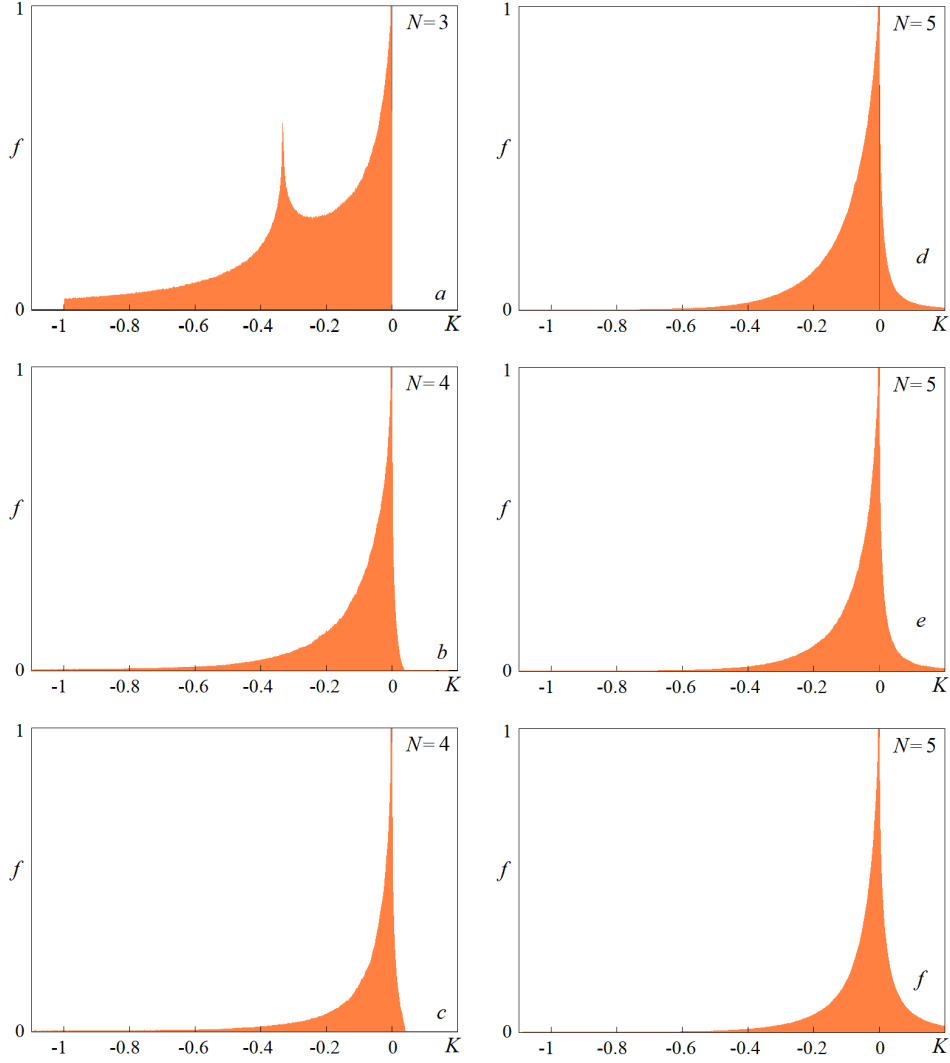}}
\label{fig4}
\caption{
Histograms of curvatures obtained from processing the data of
numerical integration of the equations of motion on a typical trajectory for
systems composed of $N$= 3 ($a)$, $N$=4 ($b, c)$ and $N$=5 ($d, e, f)$ rotators. The histograms (a),
(b), (d) correspond to the sectional curvatures in the two-dimensional directions given by
the velocity vector and the Lyapunov vector associated with the largest
Lyapunov exponent. The remaining histograms for $N$=4 and 5 refer to the
sectional curvatures in the directions given by the velocity vectors with
the second ($c$ and $e)$ and the third ($f)$ Lyapunov vectors.
}
\end{figure}

Fig. 4a shows a histogram of curvature values at points of a typical
trajectory obtained from processing the data of numerical integration of the
equations of motion over a sufficiently large time interval. From the
histogram it can be seen that the curvature values are distributed in the
range from $ - 1$ to 0, which is in accordance with formula (\ref{eq22}).

In the case of four rotators, the geodesic flow takes place on a
three-dimensional manifold. The matrix $K_{ij}$ has a size $2\times 2$.
Figure 4b shows a histogram of the values of the matrix element $K_{11}$,
which characterizes the sectional curvature of the manifold in the direction
given by the velocity vector and the Lyapunov vector corresponding to the
largest Lyapunov exponent. Fig. 4c shows the histogram for the $K_{22}$
element that is for the sectional curvature in the direction given by the
velocity vector and the second Lyapunov vector, which also corresponds to a
positive, but smaller Lyapunov exponent. As can be seen from both
histograms, the curvatures for the most part is negative, in which one can find out a correspondence with the chaotic nature of the
dynamics; however, positive curvature values also occur with
notable statistical significance.

In the case of five rotators, the motion is carried out on a four-dimensional curved manifold, and the matrix $K_{ij}$ has a size $3\times
3$. Figures 4d, e, f show the histograms of the matrix elements $K_{11}$,
$K_{22}$, $K_{33}$, which characterize the sectional curvature of the manifold
in the directions specified by the velocity vector and Lyapunov vectors,
corresponding to three positive Lyapunov exponents, namely, the largest, the
next largest and the smallest one. Again, it can be seen that the curvature
is mostly negative, but its positive values are also observed.

The fact of observation of positive values of sectional curvatures in the
last two cases $N$=4 and 5 does not necessarily indicate violation of
hyperbolicity with certainty, but closes the opportunity to use directly the
Anosov theorem concerning everywhere negative sectional curvature to justify
it.

\section{Angles between subspaces of \\ perturbation vectors}
The idea of testing hyperbolicity based on evaluating angles between
stable and unstable directions for saddle invariant sets was proposed in
\cite{27}  and after that it was applied to various situations of chaotic dynamics in a
number of specific examples of dynamical systems \cite{28,29,30}. The technique
consists in the following. A typical trajectory belonging
to the invariant set under consideration is traversed forth in time
and then back in time, and at the visited points the angles between the subspaces of vectors of small
perturbations are determined and their statistical distribution is analyzed.
If the distribution obtained does not contain angles close to zero, this
indicates the hyperbolicity of the invariant set. Contrary, if a positive
probability of zero angles is detected, then tangencies between the stable
and unstable subspaces occur, and there is no hyperbolicity.

In the case of high-dimensional systems, to identify a contracting subspace
it is more convenient to use not vectors belonging to it, but vectors
defining its orthogonal complement, whose dimension is usually small \cite{31}.
The latter are obtained from solution of the adjoint system of linearized
equations in variations \cite{31,32,33,34}.

We start with calculating the reference orbit ${\rm {\bf
x}}(t)$ integrating the equations of motion (\ref{eq8}), which we write
symbolically here as ${\rm {\bf \dot {x}}} = {\rm {\bf F}}({\rm {\bf x}},t)$, for
a sufficiently large time interval. Then we take the linearized equations for
the perturbation vectors ${\rm {\bf \dot {\tilde {x}}}} = {\rm {\bf
{F}'}}({\rm {\bf x}}(t),t){\rm {\bf \tilde {x}}}$ (equation (\ref{eq17})), in a
number equal to the dimension of the unstable subspace of interest $m$, and
integrate it along the reference trajectory ${\rm {\bf x}}(t)$ with
orthogonalization and normalization of the vectors at each step.$^{
}$\footnote{The scalar product is defined
here in the full tangent space of dimension 2$N$, as the sum of the products of the
coordinate and velocity components of the perturbation vectors, in contrast
to the previous two sections. Note that the values of the Lyapunov exponents
do not depend on the choice of one or the other definition of the scalar
product.} Next, we carry out integration in reverse time, along the same
reference trajectory, for a set of $m$ equations
${\rm {\bf \dot {u}}} = - [{\rm {\bf {F}'}}({\rm {\bf x}}(t),t)]^{T}{\rm {\bf u}}$, where the
superscript $T$ means the matrix conjugation. This gives a set of vectors
defining an orthogonal complement to the subspace of perturbations with zero
and negative Lyapunov exponents. The resulting vectors are also
orthogonalized and normalized. On the basis of these data, the angles
between the subspaces of the dimension of interest are estimated at the
points of the trajectory at each step. (The angle between two subspaces is
defined as the minimal possible angle between two vectors, one of which
belongs to one and the other to another subspace.)

\begin{figure}[htbp]
\centerline{\includegraphics[width=3.5in]{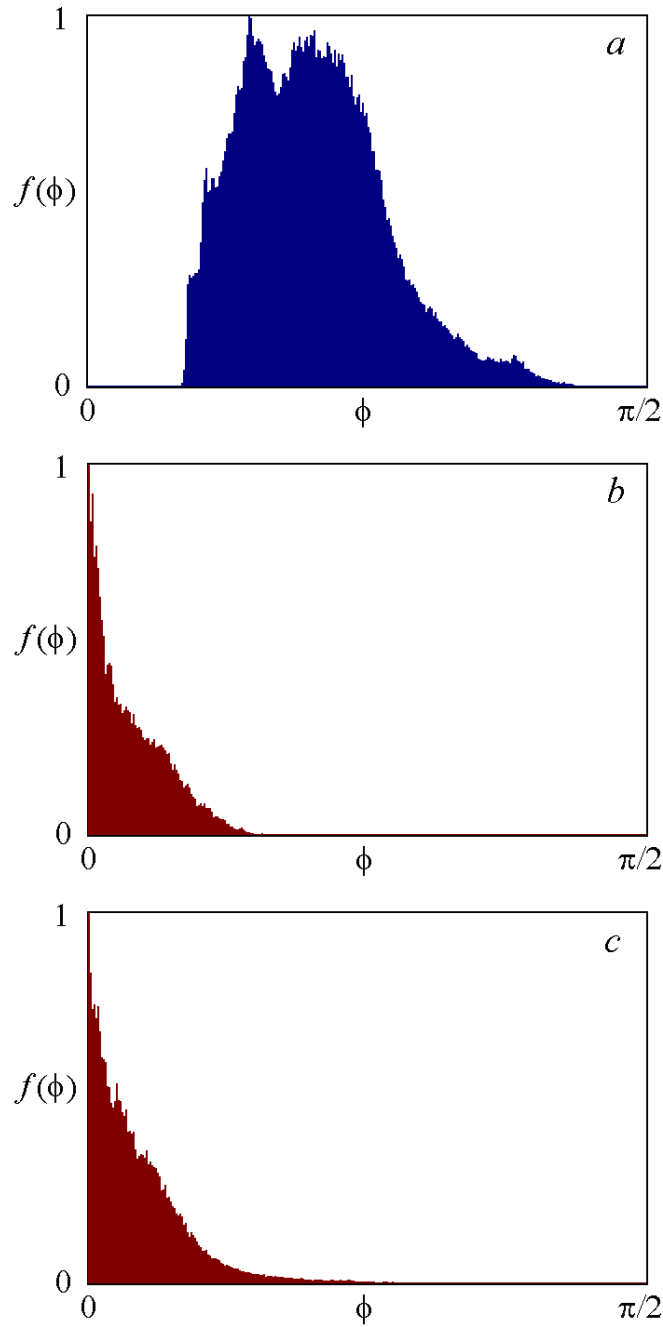}}
\label{fig5}
\caption{
Histograms for angles between unstable subspaces of
perturbation vectors of a typical trajectory and subspaces corresponding to
vectors with negative and zero Lyapunov exponents, obtained from processing
data of numerical calculations for systems based on $N$=3 ($a)$, $N$=4~($b)$ and
$N$=5~($c)$ rotators.
}
\end{figure}
Figure 5a shows the numerically obtained histogram of the distribution of
angles between a one-dimensional unstable subspace and a subspace formed by
vectors with negative and zero Lyapunov exponents, at points of a typical
trajectory for a system of three rotators with a unit velocity along the
geodesic. As seen, the distribution is obviously distant from zero values of
the angles $\phi $, that is, the test confirms the hyperbolicity of the
geodesic flow.

Figures 5b and 5c show the results of testing hyperbolicity for systems of
four and five rotators. Here are histograms of the distributions of angles
between unstable subspaces, having a dimension, respectively, two and three,
and subspaces of vectors with negative and zero Lyapunov exponents for the
case of motion along the geodesics with unit velocity in these systems.
Observe that the histograms show definitely an occurrence of angles close to
zero, which indicates the presence of tangencies of stable and unstable
subspaces and the non-hyperbolic nature of the geodesic flow.

%---------------------------------
\section{Conclusion}
The starting point of this work was the remarkable result of Hunt and MacKay
\cite{8}, who pointed out the Thurston and Weeks triple linkage mechanism \cite{7} as
an example of physically realizable mechanical system providing Anosov's
chaotic dynamics for a geodetic flow on a two-dimensional curved manifold.
The same type of dynamics takes place for a simplified version of the
system, where the condition of mechanical constraint on the elements
constituting the system has the form of vanishing the sum of the cosines of
the rotation angles \cite{12,13}. The question was whether the dynamics of
systems built by this sample with a greater number of elements would have
similar properties. In the present work, I give a negative answer, at least
for the models where the constraint condition is written as equality to zero
of the sum of the cosines for the revolution angles of the rotators.

Namely, numerical calculations show that for the number of elements four and
five, when the dynamics correspond to geodesic flows on three-dimensional
and four-dimensional manifolds, the conditions of negative sectional
curvature, which would be sufficient for realization of the Anosov dynamics,
do not hold. Also, on the basis of numerical calculations, it was found that
for systems of four and five elements, the hyperbolicity condition requiring
the transversality of the intersection of subspaces associated with positive
Lyapunov exponents and with non-positive exponents is not satisfied. These
results contrast with the case of three rotators corresponding to the
geodesic flow on a two-dimensional manifold, the Schwarz surface, where the
calculations carried out using the same algorithms confirm belonging to the
class of Anosov systems and the hyperbolicity condition is satisfied in the
sense of the absence of tangencies between the subspaces of the perturbation
vectors.

Nevertheless, the dynamics in all the considered cases is chaotic, as
evidenced by the observed waveforms, calculations of Lyapunov exponents, and
power spectra of the motions generated by the systems.

For a system of three elements, an approach was previously developed that
allows one to arrange the Anosov dynamics on attractor in a system that is
an electronic analog, a chaos generator with spectral characteristics
satisfactory from the point of view of radiophysical applications \cite{12,13,14}.
Similarly, it is possible to implement electronic analogs for the systems
composed of four, five or more elements, which will generate hyperchaos,
the chaotic dynamics with two or more positive Lyapunov exponents.

Returning to the mechanical systems in the form of rotators with constraints
determined by algebraic equations, it can be noted that the question of a
possible occurrence of Anosov's dynamics for geodesic flows on
manifolds of dimension three and higher in such systems cannot be considered
completely closed. For example, a possibility remains unexplored of using
various distributions of masses among the elements constituting the system.
\vspace{3ex}

{\it This work was supported by Grant no. 15-12-20035 of the Russian Science Foundation.
The author thanks P.V.~Kuptsov for help in calculating the
angles between the subspaces of the perturbation vectors, the results of
which are presented in Section 4.}
\vspace{3ex}
%----------------------------------------------

\noindent{\bf{Appendix}}
\vspace{3ex}

\noindent
The flat linkage shown in Fig. 1b can be interpreted as four
rotators that revolve around axes located at the vertices of a square with
coordinates $(\pm 1,\,\,0)$ and $(0,\,\,\pm 1)$.
The imposed constraint is due to the connection of the rotators by the rods
of length $R$, attached to them by hinges at a distance $\varepsilon $ from the
axes and also attached to the hinges placed at the vertices of a square on a moving disk,
at a distance $\rho $ from its center. Let $x$ and $y$ be coordinates of the
center of the disk, and $\alpha $ is the rotation angle of the disk. The
rotation angles of the rotators are counted from vertical or horizontal coordinate
axes, as shown in the figure, so that the coordinates of the end joints are
\[
\begin{array}{l}
 \,(1 - \varepsilon \cos \theta _0 ,\,\,\varepsilon \sin \theta _0
),\,\,(\,\varepsilon \sin \theta _1 ,\,1 + \varepsilon \cos \theta _1
),\,\,\, \\
 ( - 1 + \varepsilon \cos \theta _2 ,\,\, - \varepsilon \sin \theta _2
),\,\,( - \varepsilon \sin \theta _3 ,\, - 1 - \varepsilon \cos \theta _3 ).
\\
 \end{array}
\]
The coordinates of the four hinges located on the disk are
\[
\begin{array}{l}
 (x + \rho \cos \alpha ,\,\,y + \rho \sin \alpha ),\,\,\,(x - \rho \sin
\alpha ,\,\,y + \rho \cos \alpha ),\, \\
 \,(x - \rho \cos \alpha ,\,\,y - \rho \sin \alpha ),\,\,(x + \rho \sin
\alpha ,\,\,y - \rho \cos \alpha ). \\
 \end{array}
\]

An instantaneous configuration of the system is given by the angular
variables $\theta _0 ,\,\theta _1 ,\,\theta _2 ,\,\theta _3 $, of which only
three are independent due to the imposed mechanical constraint. Thus, the
configuration space is a three-dimensional manifold. The equations that
define the mechanical constraint are
\[
\begin{array}{l}
 (1 - \varepsilon \cos \theta _0 - x - \rho \cos \alpha )^2 + (\,\varepsilon
\sin \theta _0 - y - \rho \sin \alpha )^2 = R^2,\,\,\, \\
 (\,\varepsilon \sin \theta _1 - x + \rho \sin \alpha )^2 + (\,1 +
\varepsilon \cos \theta _1 - y - \rho \cos \alpha )^2 = R^2, \\
 ( - 1 + \varepsilon \cos \theta _2 - x + \rho \cos \alpha )^2 + (\, -
\varepsilon \sin \theta _2 - y + \rho \sin \alpha )^2 = R^2, \\
 ( - \varepsilon \sin \theta _3 - x - \rho \sin \alpha )^2 + (\, - 1 -
\varepsilon \cos \theta _3 - y + \rho \cos \alpha )^2 = R^2. \\
 \end{array}
\]

Let us turn now to consideration of the asymptotic case when $\varepsilon < <
R$, $\rho \sim \varepsilon ,\,\,(x,\,y)\sim \varepsilon $, and $R$=1. Taking
into account the terms of the first order in the Taylor expansions, we have
\[
\begin{array}{l}
 - \varepsilon \cos \theta _0 - x - \rho \cos \alpha = 0,\,\,\, -
\varepsilon \cos \theta _1 + y + \rho \cos \alpha = 0, \\
 - \varepsilon \cos \theta _2 + x - \rho \cos \alpha = 0,\,\,\, -
\varepsilon \cos \theta _3 - y + \rho \cos \alpha = 0. \\
 \end{array}
\]
Summating the equations, we get
\[
\cos \theta _0 + \cos \theta _1 + \cos \theta _2 + \cos \theta _3 = 0.
\]

Assuming that the massive elements of the system are only the rotators,
each characterized by a unit moment of inertia, for the kinetic energy
we have to write $T = \textstyle{1 \over 2}(\dot {\theta }_0^2 + \dot
{\theta }_1^2 + \dot {\theta }_2^2 + \dot {\theta }_3^2 )$. For our system,
this expression is just the Lagrange function. From the conditional
extremum of the action functional $\int {(T - \Lambda F)dt} $ with taking
into account the imposed constraint, we obtain finally the
equation of motion in the form (\ref{eq7}) or (\ref{eq8}) for $N$=4.

%----------------------------------------------

\end{document}